# Structure-dependent microwave dielectric properties of $(1-x)\mathrm{La}(\mathrm{Mg}_{1/2}\mathrm{Ti}_{1/2})\mathrm{O}_3 - x\mathrm{La}_{2/3}\mathrm{TiO}_3$ ceramics

Andrei N. Salak[a)], Dmitry D. Khalyavin, Pedro Q. Mantas, Ana M.R. Senos
and Victor M. Ferreira

*Department of Ceramics and Glass Engineering/CICECO, University of Aveiro, 3810-193 Aveiro, Portugal*

**Abstract**

Structure and dielectric characterization were performed on the $(1-x)\mathrm{La}(\mathrm{Mg}_{1/2}\mathrm{Ti}_{1/2})\mathrm{O}_3 - x\mathrm{La}_{2/3}\mathrm{TiO}_3$ perovskite ceramics with the actual composition $0 \leq x \leq 0.52$. The sequence of structure transformations was detected as $x$ is increased: $P2_1/n$ ($0 \leq x < 0.3$) → $Pnma$ ($0.3 \leq x < 0.37$) → $Imma$ ($0.37 \leq x < 0.4$) → $I2/a$ ($0.4 \leq x \leq 0.49$) → $R\bar{3}c$ ($0.49 < x \leq 0.52$). The structure evolution from the tilt configuration $a^+b^-b^-$ to $a^-a^-a^-$ are considered in terms of competing the repulsive energy and the Madelung energy. Relative permittivity of the ceramics was measured as a function of temperature and LT content. Temperature coefficient of the resonant frequency was found to pass zero-value at the narrow compositional range between $x=0.49$ and $0.52$, where the discontinuous $I2/a \to R\bar{3}c$ crossover occurs. Temperature and compositional variation of these fundamental microwave dielectric parameters are discussed in respect to type of the phase transitions (continuous/discontinuous) between the structures.

**1. Introduction**

Due to almost mutually exclusive demands to low-loss materials for microwave applications, namely both high and temperature-stable dielectric permittivity, these cannot be ferroelectric [1]. At the same time, most of the microwave materials have been found in structures related to perovskite, pyrochlore, tungsten-bronze and some others commonly considered as ferroelectricity-bearing. All these have linked oxygen octahedra as a structure frame. Provided that the so-called "ferroelectrically active" cations, such as $\mathrm{Ti}^{4+}$, $\mathrm{Nb}^{5+}$ and $\mathrm{Ta}^{5+}$, are within the octahedra, in weak fields these act as the polar units making high enough the permittivity of the respective compositions.

Whatever the crystal structure of $\mathrm{Ln}_2\mathrm{O}_3$-$\mathrm{TiO}_2$-based compositions (Ln: lanthanide), these generally exhibit promising dielectric properties at GHz range [1-3]. To adjust the fundamental dielectric parameters, these were often made to be multi-component and their exact crystal structure was not well understood [1]. One of few, the systematic investigation of phase relations, crystal structure and related dielectric properties has been recently performed in the $\mathrm{La}_2\mathrm{O}_3$-$\mathrm{CaO}$-$\mathrm{MgO}$-$\mathrm{TiO}_2$ system [4].

It was shown that $\mathrm{Ln}(\mathrm{B}^{2+}_{1/2}\mathrm{Ti}_{1/2})\mathrm{O}_3$ perovskites, presenting oxygen octahedra tilting and near perfect B-site cation ordering, have promising microwave properties, particularly a low dielectric loss [5]. In this family, $\mathrm{La}(\mathrm{Mg}_{1/2}\mathrm{Ti}_{1/2})\mathrm{O}_3$ (LMT) was found to give the highest values of both relative permittivity ($\varepsilon_r = 27.4$) and quality factor ($Q = 16110$ at the resonant frequency, $f_0 = 7.1$ GHz) [6]. However, the relatively high magnitude of its temperature coefficient of the resonant frequency ($\tau_f = -81$ ppm K$^{-1}$) does this compound in itself out of use in microwave devices. Tuning of $\tau_f$ to near-zero values may be achieved by the formation of solid solutions with compounds having the thermostability coefficients of opposite sign. It has been successfully

---

[a)] *Author to whom any correspondence should be addressed*
*E-mail address:* salak@cv.ua.pt (A.N. Salak)

realized in LMT-based solid solutions with the earth alkaline titanates: $BaTiO_3$ [6], $SrTiO_3$ [7] and $CaTiO_3$ [8].

Similarly, the perovskite solid solutions between LMT and $La_{2/3}TiO_3$ (LT) were attempted [4,9]. The orthorhombic structure of LT was early considered as "double perovskite" because its parameters are $a \approx a_p$, $b \approx a_p$ and $c \approx 2a_p$ ($a_p$ – parameter of cubic perovskite) [10]. The $c$-axis length is doubled due to the ordered arrangement of the $La^{3+}$ cations and vacancies in $A$-site of the perovskite lattice. It has been recently revealed that the crystal structure of LT is orthorhombic $Cmmm$ with antiphase tilting of the $TiO_6$ octahedra around an axis perpendicular to the direction of the $A$-site ordering [11]. Owing to the high vacancy content, the perovskite phase of stoichiometric $La_{2/3}TiO_3$ does not occur. To be stabilized it needs a partial reduction of $Ti^{4+}$ or a partial heterovalent substitution at $A$- or $B$-sites [12,13]. Therefore the dielectric properties of LT were studied mainly in the solid solutions with other oxide compounds [10,14]. It has been found that the (1-$x$)LMT-$x$LT system presents a limited solubility (0≤$x$≤0.5) [4,9]. Non-linear volume cell dependence and peculiar variation of the fundamental dielectric characteristics, namely $\varepsilon_r$, $Q \times f_0$ and $\tau_f$, were observed as a function of $x$ [4]. It was previously reported the structure of particular compositions only [9]. Vanderah *et al.* also suggested that systematic crystal structure changes take place in the system [4] but the true pattern of structure evolution in the (1-$x$)LMT-$x$LT system has not been clarified yet.

It is known that, besides crystal chemistry of "microwave perovskites", the particular features of their structure, such as cation ordering and oxygen octahedra tilting, affect notably the fundamental dielectric parameters: $\varepsilon_r$, $Q$ and $\tau_f$ [15,16]. At the same time, the effect of cation vacancies has not been systematically studied yet in low-loss perovskites. It seems the observed limited solubility and the peculiarities of dielectric behavior versus composition could be explained in terms of the crystal chemistry and structure of LMT-LT. Therefore, a detailed investigation of the crystal structure sequence in the system certainly deserves attention.

This work reports on the structure sequence and related microwave dielectric properties of the (1-$x$)LMT-$x$LT perovskite ceramics with the actual composition 0≤$x$≤0.52. Variation of relative permittivity and temperature coefficient of the resonant frequency as a function of composition is discussed in terms of phase transitions altering the oxygen octahedra tilting in the system.

## 2. Experimental

(1-$x$)LMT-$x$LT solid solutions with the nominal compositions 0≤$x$≤0.55 were processed from powders obtained by a citrate-based chemical route [17]. Calcined powders were uniaxially pressed into disks of 10 mm in diameter and about 1 mm in thickness and sintered in air at temperatures ranging from 1720 to 1820 K for 2 hours. The heating and cooling rates were 10 K min$^{-1}$ in every case. For microwave dielectric measurements, cylindrical samples of 10 mm in diameter and 8-10 mm length were isostatically pressed and sintered under the same conditions.

Phase analysis of the ceramics and crystal structure determination were performed by X-ray diffraction (XRD) from the powders of the samples ground (Rigaku D/MAX-B diffractometer, Cu K$\alpha$ radiation, tube power 40 kV, 30 mA; graphite monochromator, receiving slit 0.15 mm). The room temperature XRD data were collected over the angular range 10<2$\theta$<120° with the step 0.02° and exposition 10 s/step. The temperature XRD experiments were carried out for the composition $x$=0.45 using a Philips X'Pert MPD diffractometer (Cu K$\alpha$ radiation, tube power 40 kV, 30 mA; graphite monochromator, receiving slit 0.15 mm). The full spectrum was recorded at 405 K (2$\theta$ range 10-120°, step 0.02°, 5 s/step), whereas at intermediate temperatures only selected reflections were examined. The Rietveld refinement of the obtained spectra was performed using the FullProf suite [18].

The samples for radio frequency dielectric measurements were polished to form disks with a thickness of 0.4-0.5 mm, electroded with platinum paste and annealed at 1000 K. Dielectric permittivity and loss tangent were measured as a function of temperature at a frequency range $10^2$ - $10^6$ Hz, using a Precision LCR meter (HP 4284A). Measurements were performed over the interval 300-500 K on cooling with a rate of 1 K min$^{-1}$. The room temperature permittivity, $Q$-factor and resonant frequency of the samples at microwave frequency range were estimated by an adaptation of the Hakki-Coleman method [19,20] using a 10 MHz - 20 GHz Scalar Analyser (IFR 6823).

## 3. Results and Discussion

Analysis of the XRD spectra of the (1-$x$)LMT-$x$LT system revealed that a single perovskite phase is formed at the range $x \leq 0.45$. Above this composition the second phase, namely $La_2Ti_2O_7$, was found to appear[1] (Fig. 1). Nevertheless, it will be shown below, the crystal structure symmetry of the ceramics with the nominal composition $x$=0.55 is different from that of the ceramics with $x$=0.5. This points to the fact that the solubility in the (1-$x$)LMT-$x$LT system occurs also above $x$=0.45 in spite of a presence of the second phase. The actual LT content in those solid solutions was estimated from the Rietveld refinement (Table I). Hereinafter the actual compositions only are referred.

*Table I. Structural parameters of the (1-x)LMT-xLT system obtained from the refinement of the room temperature XRD spectra*

| $x$ | Space group | $a$ (Å) | $b$ (Å) | $c$ (Å) | $\beta$ (°) | $V$ (Å$^3$) | $Z$ |
|---|---|---|---|---|---|---|---|
| 0 | $P2_1/n$ | 5.5644(10) | 5.5742(10) | 7.8665(20) | 90.03(1) | 243.998(10) | 4 |
| 0.1 | $P2_1/n$ | 5.5555(11) | 5.5595(10) | 7.8632(19) | 90.06(2) | 242.865(20) | 4 |
| 0.2 | $P2_1/n$ | 5.5484(10) | 5.5620(10) | 7.8478(20) | 90.06(1) | 242.186(10) | 4 |
| 0.3 | $Pnma$ | 5.5596(13) | 7.8316(20) | 5.5392(11) | | 241.182(10) | 4 |
| 0.37 | $Imma$ | 7.8234(19) | 5.5592(12) | 5.5334(13) | | 240.656(10) | 4 |
| 0.4 | $I2/a$ | 7.8199(19) | 5.5592(11) | 5.5305(13) | 90.08(1) | 240.427(9) | 4 |
| 0.45 | $I2/a$ | 7.8145(16) | 5.5555(10) | 5.5270(11) | 90.14(1) | 239.945(8) | 4 |
| 0.49 (0.50)** | $I2/a$ | 7.8113(30) | 5.5488(17) | 5.5219(18) | 90.24(1) | 239.334(14) | 4 |
| 0.52 (0.55)** | $R\bar{3}c$ | 5.5392(13) | | 13.4925(4) | | 358.519(16) | 6 |

*The data for LMT ($x$=0) were taken from Ref. [24].
**The numbers in brackets denote the nominal composition $x$ meant when preparing the solid solution.

Symmetry of the crystal structure of LMT was a matter of discussion in several works [21-23]. Three different space groups were proposed for this perovskite composition. The structure of LMT was earlier reported [21] to be a cubic one with $Pa3$ space group. Later the structure was re-refined and lower-symmetry space groups, *viz.* orthorhombic $Pnma$ and monoclinic $P2_1/n$ ($P2_1/c$ in standard setting) have been proposed by Meden and Ceh [22] and Lee *et al*. [23], respectively. Avdeev *et al*. [24] have recently confirmed that at room temperature LMT has the monoclinic symmetry with $P2_1/n$ space group, which allows to account for both rock salt type ordering between $Mg^{2+}$ and $Ti^{4+}$ ions and $a^+b^-b^-$ octahedral tilting. Therefore, this space group was chosen for the refinement of diffraction data for the LMT-rich compositions.

---

[1] According to Ref. [4], there should be also $MgTiO_3$, but the amount of this phase seemed to be too small to be detected.

Dilution of LMT with LT results in a gradual decreasing intensity of the (*1/2 1/2 1/2*) superstructure reflection associated with Mg/Ti cation ordering. The long-range ordering in the (1-*x*)LMT-*x*LT ceramics disappears when $0.2<x<0.3$ (Fig. 1). However, the superstructure reflections originated from antiparallel displacements of *A*-site cations 1/2(*eeo, eoe, oee*; *o*-odd, *e*-even), in-phase 1/2(*ooe, ooe, eoo*) and antiphase 1/2(*ooo*; $h+k+l>3$) octahedral tilting [25] are still clearly observed for the composition *x*=0.3 (inset in Fig. 1). This means that the change of the symmetry from monoclinic $P2_1/n$ ($a^+b^-b^-$) to orthorhombic $Pnma$ ($a^+b^-b^-$) [26,27] takes place at the range $0.2<x<0.3$. The refinement of XRD data for ceramics with *x*=0.3 was successful with correct description of all the observed intensities. In the case of the $a^+b^-b^-$ tilt system, the perovskite primitive unit cell (containing one formula unit $ABO_3$) is characterized by the pseudomonoclinic metric with $a_p=c_p\neq b_p$ and $\alpha_p=\gamma_p=90°\neq\beta_p$ (where $a_p$, $c_p$, $b_p$ and $\alpha_p$, $\gamma_p$, $\beta_p$ are the parameters of the primitive unit cell). This results in a splitting of both (*hkl, k=l=0*) and (*hkl, h=k=l*) fundamental reflections into doublets with the intensity ratio between reflections belonging to a single multiplet being 2:1 and 1:1, respectively (Fig. 2)[2].

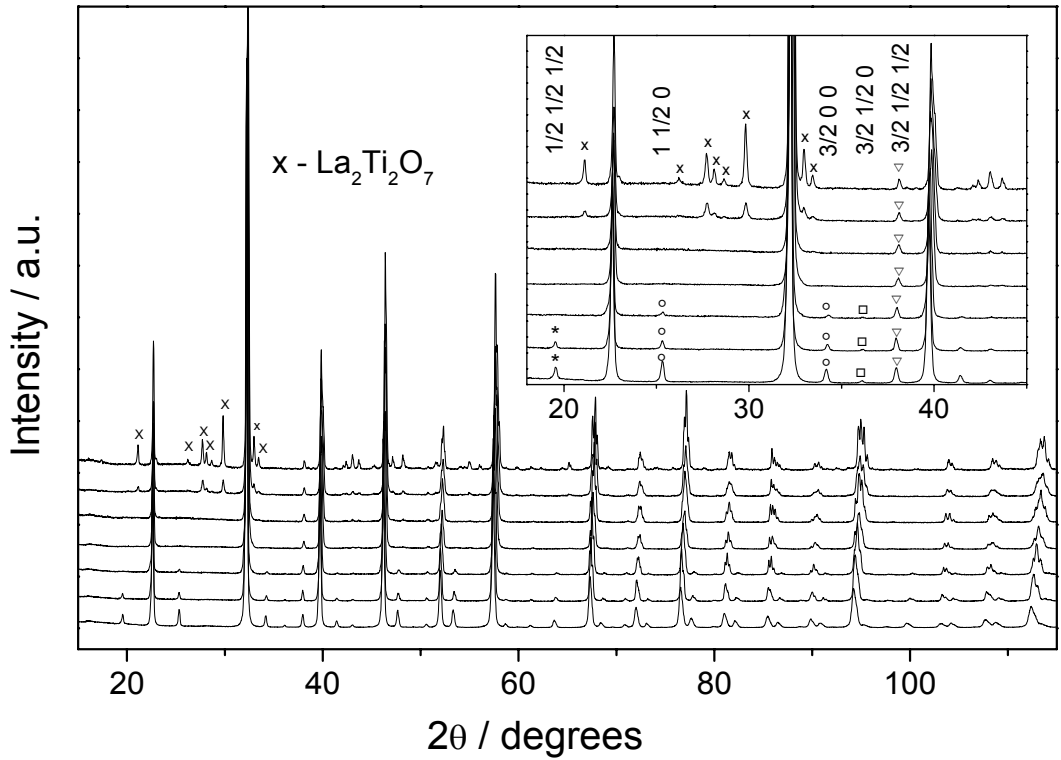

***Fig. 1.*** *XRD pattern of (1-x)LMT-xLT with the nominal composition x=0 (bottom), 0.2, 0.3, 0.37, 0.45, 0.5 and 0.55 (top). Inset shows superstructure reflections indicating Mg/Ti ordering (∗), antiparallel La displacements (○), in-phase octahedra tilting (□) and antiphase octahedra tilting (▽).*

Further increase of LT content to *x*=0.37 yields disappearance of the superstructure reflections associated with both in-phase octahedral tilting and antiparallel displacements of *A*-site cations, whereas the splitting of the fundamental reflections does not change qualitatively (Fig. 2). It reflects the vanishing of in-phase rotations of the oxygen octahedra and hence a change of the symmetry from orthorhombic *Pnma* ($a^+b^-b^-$) to orthorhombic *Imma* ($a^0b^-b^-$) [25,26].

---

[2] $P2_1/n$ space group allows the $a_p$ and $c_p$ parameters are not to be equal each other.

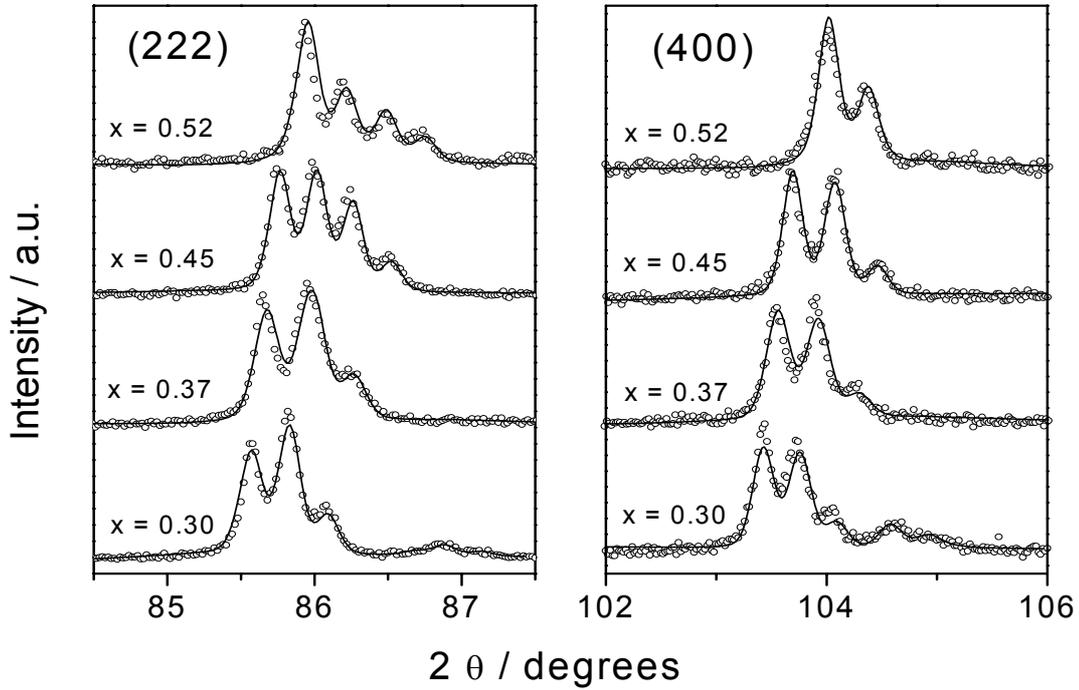

***Fig. 2.*** *Room temperature XRD patterns at the regions of (222) and (400) fundamental reflections for selected compositions of (1-x)LMT-xLT refined by Rietveld method (solid line).*

The XRD spectra of the ceramics with $0.4 \leq x \leq 0.49$ show the presence of the superstructure reflections related to antiphase tilting (inset in Fig. 1) and a splitting of the (*hkl*, *h*=*k*=*l*) and (*hkl*, *k*=*l*=*0*) fundamental multiplets into a triplet and a doublet, respectively (Fig. 2). This splitting suggests the situation, when all three angles of the primitive perovskite unit cell are not equal to 90° ($\alpha_p$, $\beta_p$, $\gamma_p \neq 90°$) and two of the angles and two of the cell parameters are equal to each other (i.e. $\alpha_p=\gamma_p\neq\beta_p$ and $a_p=c_p\neq b_p$). In that case the primitive perovskite unit cell is characterized by a set of four independent parameters. Such a pseudo-triclinic primitive unit cell can be transformed into a monoclinic one (see Appendix) and the crystal structure is then described by the *I2/a* space group (*C2/c* in standard setting), corresponding to the $a^-b^-b^-$ tilt system [25,26]. This space group was successfully used for the refinement procedure of XRD patterns for the compositions with *x*=0.4-0.49.

Increasing the actual LT content in (1-*x*)LMT-*x*LT up to *x*=0.52 results in one more phase transition. This conclusion is drawn again from the analysis of the (*hkl*, *k*=*l*=*0*) and (*hkl*, *h*=*k*=*l*) fundamental reflections. It was revealed that the first multiplet type is a singlet and the second one is a doublet with the intensity ratio of 3:1 between the doublet components. The observation indicates that the primitive unit cell is characterized by the metric with $a_p=b_p=c_p$ and $\alpha_p=\beta_p=\gamma_p\neq 90°$. This fact in combination with the 1/2(*ooo*, *h*+*k*+*l*>3) superstructure reflections observed (inset in Fig. 1) suggests that the composition with *x*=0.52 has the rhombohedral symmetry with the $R\bar{3}c$ space group. This symmetry is a result of antiphase rotations of the oxygen octahedra around the three crystallographic axes of the primitive unit cell with equivalent rotation angles and hence corresponds to the $a^-a^-a^-$ tilt system [25,26]. Thus, the refinement for the main phase was performed using $R\bar{3}c$ space group; the second $La_2Ti_2O_7$ phase (ICSD Collection code 72433) was introduced into the calculation with the $P2_1$ space group.

Based on the refinement results (Table I) and the unit cell relations summarized in Appendix, the parameters of the primitive perovskite unit cell have been calculated. Their composition evolution is presented in Figure 3. As seen, the parameters vary gradually and there is the only discontinuity at the $I2/a \to R\bar{3}c$ phase transformation. It is consistent with the results of the group-theoretical analysis performed by Howard and Stokes [26]. According to them, the $Pnma \to Imma$ and $Imma \to I2/a$ phase transformation are allowed by Landau theory to be continuous, whereas the $I2/a \to R\bar{3}c$ transition is required to be discontinuous. With the aid of the *ISOTROPY* software [28], we extended the symmetry analysis to the case of the $P2_1/n \to Pnma$ phase transition. It was found the rock salt type cation ordering is associated with one-dimensional irreducible representation $\Gamma_3^+(k=0,0,0)$ of the $Pnma$ space group. The phase transition mediated by this representation is allowed to be continuous in renormalization-group theory. Indeed, the anomaly-less monoclinic $P2_1/n$ to orthorhombic $Pnma$ phase transformation is observed when $x$ is increased from 0.2 to 0.3 (Fig. 3).

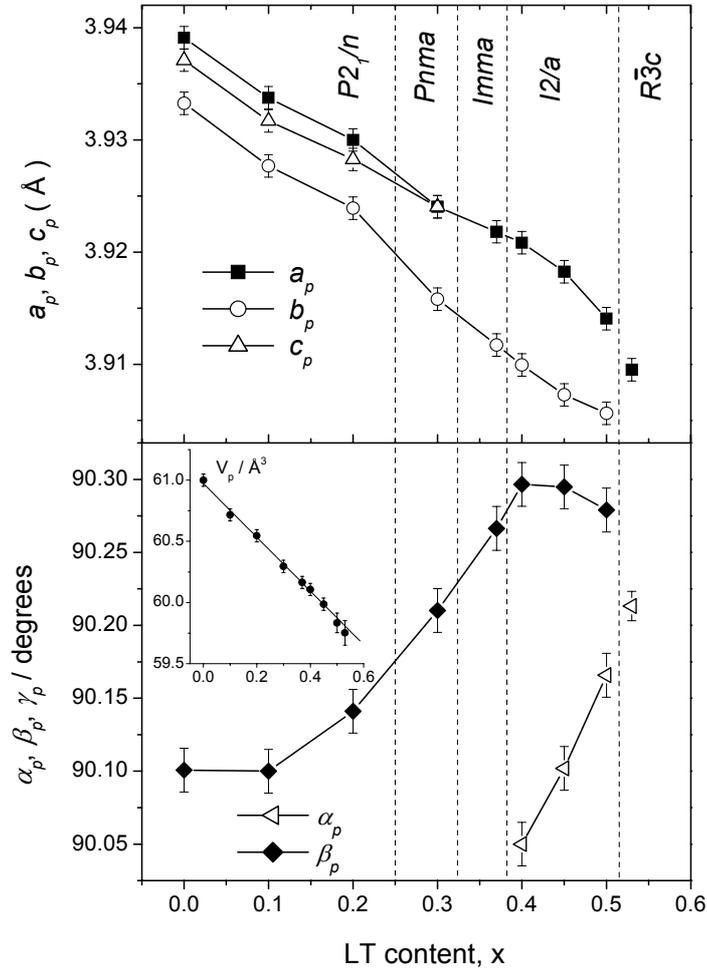

*Fig. 3.* Primitive perovskite unit cell parameters ($a_p$, $b_p$, $c_p$ and $\alpha_p$, $\beta_p$, $\gamma_p$) as a function of LT content in the (1-x)LMT-xLT system. **Inset:** compositional variation of primitive perovskite unit cell volume.

The structure evolution in the LMT-LT system, from the $a^+b^-b^-$ to $a^-a^-a^-$ tilt configurations, can be understood in terms of the size effect at $Mg^{2+} \to Ti^{4+}$ substitution and the influence of vacancies in the La-position. According to Woodward [29], the magnitudes of lattice energy associated with $a^+b^-b^-$ and $a^-a^-a^-$ tilt systems are very close. The $a^+b^-b^-$ tilt system is favored in respect of the repulsive energy because this tilt system allows $A$-site cations to be shifted. At the same time the $a^-a^-a^-$ tilt system requires $A$-site cations to remain in their highly symmetrical positions and is favored in respect of the Madelung energy. In the case of the $a^+b^-b^-$ tilt system a reduction of the repulsive energy due to a shift of $A$-site cations is the greater, the tilt angles are larger. Therefore this tilt system is usually realized in perovskites with a small tolerance factor. In the LMT-LT system, the substitution of $Mg^{2+}$ cation (0.88 Å) with smaller $Ti^{4+}$ (0.74 Å) results in an increase of the tolerance factor. Besides, an amount of the vacancies in the La-position is increased with $x$; thereby a number of the shifted cations is decreased. Thus, both the size effect and the $A$-site vacancies suppress the factors stabilizing the $a^+b^-b^-$ tilt system and this is probably reason why the $a^-a^-a^-$ tilt system becomes more stable at higher LT content.

Dielectric characteristics of the (1-$x$)LMT-$x$LT ceramics ($0 \leq x \leq 0.49$) measured at microwave frequency range are summarized in Table II. It should be noted that the observed reduction of the $Q \times f_0$ magnitude with composition is almost one order of magnitude within the narrow composition range siding with LMT. Some authors attributed such huge decrease of the quality-factor to the loss of the B-site cation ordering [30]. However, it seems to be not the only reason. In the system under study, the Mg/Ti ordered arrangement was proved to be at least in the ceramics with LT content as high as 20 mol% where the main increase of dielectric loss occurs. Previous works on the (1-$x$)LMT-$x$LT system also suggested that the dependence of $Q \times f_0$ on $x$ is not evident [4,9]. Hereinafter we concern mainly the dielectric permittivity and thermal coefficient of the resonant frequency. Likely origins of the observed behavior of microwave dielectric loss will be considered elsewhere.

***Table II**. Microwave dielectric properties in the (1-x)LMT-xLT system*

| $x$ | Relative density (%) | $\varepsilon_r$ | $f_0$ (GHz) | $Q \times f_0$ (GHz) |
|---|---|---|---|---|
| 0* | 97 | 27.6 | 7.10 | 114312 |
| 0.1 | 95 | 26.5 | 8.30 | 30577 |
| 0.2 | 90 | 27.5 | 7.89 | 16595 |
| 0.3 | 92 | 29.3 | 8.01 | 6520 |
| 0.37 | 91 | 33.2 | 8.29 | 8563 |
| 0.4 | 91 | 35.1 | 7.96 | 6675 |
| 0.45 | 92 | 39.7 | 7.57 | 5826 |
| 0.49 | 96 | 46.5 | 6.20 | 8307 |

*The data for LMT ($x$=0) were taken from Ref. [7].

Relative permittivity of the ceramics was also measured as a function of temperature at radio frequency range (1 MHz). As $\varepsilon_r$(T) varied almost linearly within 300-360 K, the thermal coefficient of capacitance ($\tau_C$) could be readily evaluated. The values of $\tau_f$ were then determined by using the relation $\tau_f = \frac{1}{2}[\tau_C + \alpha_L]$ [31]. The linear thermal-expansion coefficient ($\alpha_L$) was assumed to be 10 ppm K$^{-1}$ as its value is ordinarily in the range 9-12 ppm K$^{-1}$ for most perovskites [32].

The relevant dielectric parameters, obtained by different methods on the (1-$x$)LMT-$x$LT ceramics are shown in Figure 4 as a function of $x$. It is evident that $\varepsilon_r$ measured at GHz range and at 1 MHz show the same behavior with increasing LT content. Difference between $\varepsilon_r$ values for each composition was found to be less than 4%. Besides, the observed behavior of both permittivity and thermal coefficient of the resonant frequency are in good agreement with the results reported by Vanderah *et al.* [4]. Unlike the $Q \times f_0$ value, the $\varepsilon_r$ and $\tau_f$ values do not exhibit any drastic change while $0 \leq x \leq 0.45$. The structure transformations which occur at this range seem not to affect notably the compositional dependence of these parameters.

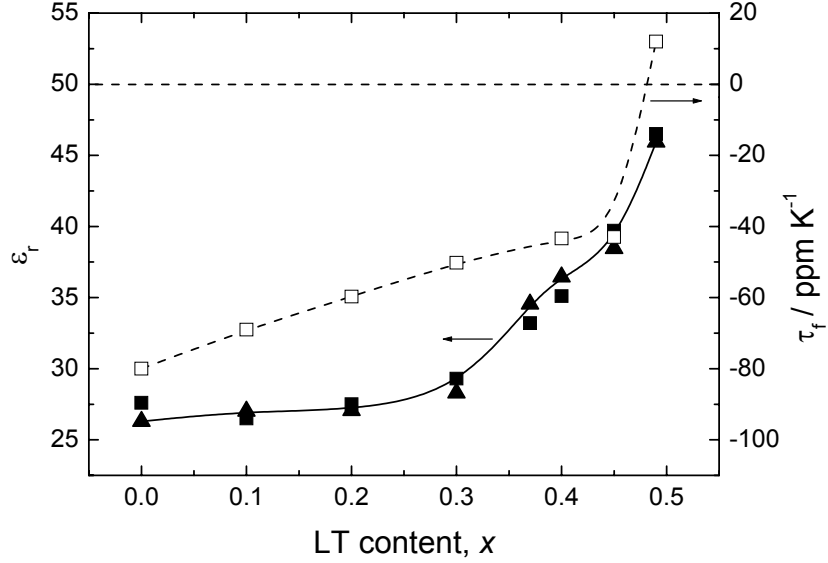

***Fig. 4.*** *Relative permittivity ($\varepsilon_r$, solid symbols) and temperature coefficient of the resonant frequency ($\tau_f$, open symbols) for the (1-x)LMT-xLT system measured at microwave (■) and radio frequencies (▲).*

It has been found that the temperature coefficient of the dielectric permittivity ($\tau_\varepsilon$) of low dielectric constant perovskites is mainly driven by the type of their crystal structure distortion in terms of the oxygen octahedra tilting [16,32]. The relationship between $\tau_\varepsilon$ and crystal structure of the perovskites was clearly drawn by Reaney *et al.* [16]. They have represented this "$\tau_\varepsilon$ versus crystal structure" relation in terms of tolerance factor (*t*) which reflects deviations from the ideal perovskite structure. According to this empirical relationship, the phase transitions involving the onset of (another type of) octahedral tilting result in alter of $\tau_\varepsilon$ and therefore $\tau_f$ of microwave perovskites. A number of perovskite compositions follow the $\tau_\varepsilon(t)$ relation; however, at the considered range of the *t*-factor variation ($0.92 \leq t \leq 1.06$) a series of structure transitions may occur. Most of those would involve a re-arrangement of the metal-oxygen network. Apparently, not all the transitions could have a notable effect on the $\tau_\varepsilon$ / $\tau_f$ behaviors. It seems reasonable therefore to consider structure transformations in respect to the type of transition. One can expect a change in the relative permittivity behavior and thereby of the magnitude of both $\tau_\varepsilon$ and $\tau_f$, if the transition between tilted structures is discontinuous.

It the (1-*x*)LMT-*x*LT system, when *x* varies from 0 to 0.45, the sequence of structure transformations, $P2_1/n \rightarrow Pnma \rightarrow Imma \rightarrow I2/a$, is observed. All these are of the second order type (continuous). Indeed, any significant changes in the compositional dependence of both $\varepsilon_r$ and $\tau_f$ are not detected at this range (see Fig. 4). At the same time, these parameters become more sensitive to the composition change as soon as *x*>0.45. It was also observed that $\tau_f$ passes zero-value at the narrow range between *x*=0.49 and 0.52, where the discontinuous $I2/a \rightarrow R\bar{3}c$ crossover was revealed. As increasing *x* (and therefore, *t*-factor) and increasing temperature can yield a similar effect on $\tau_\varepsilon$ / $\tau_f$ [16], relative permittivity of the ceramics with $x \geq 0.45$ was measured at an extended temperature range.

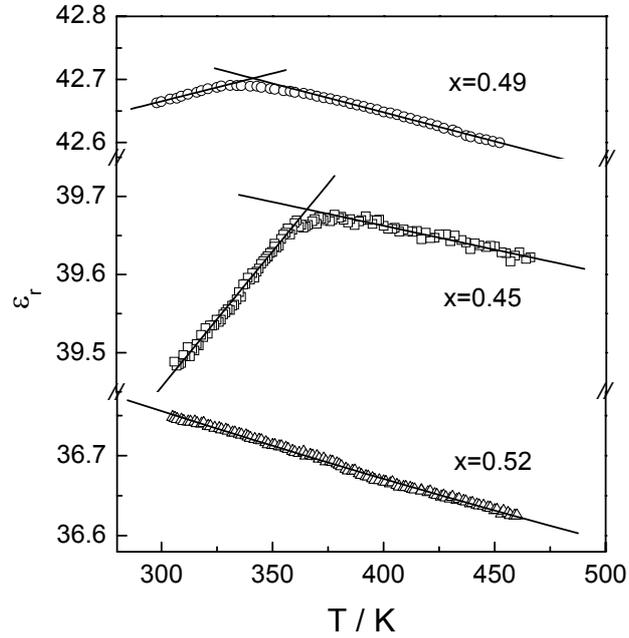

*Fig. 5.* Temperature variation of relative permittivity ($\varepsilon_r$) for the (1-x)LMT-xLT ceramics measured at 1 MHz (symbols and solid lines represent experimental data and linear fits, respectively).

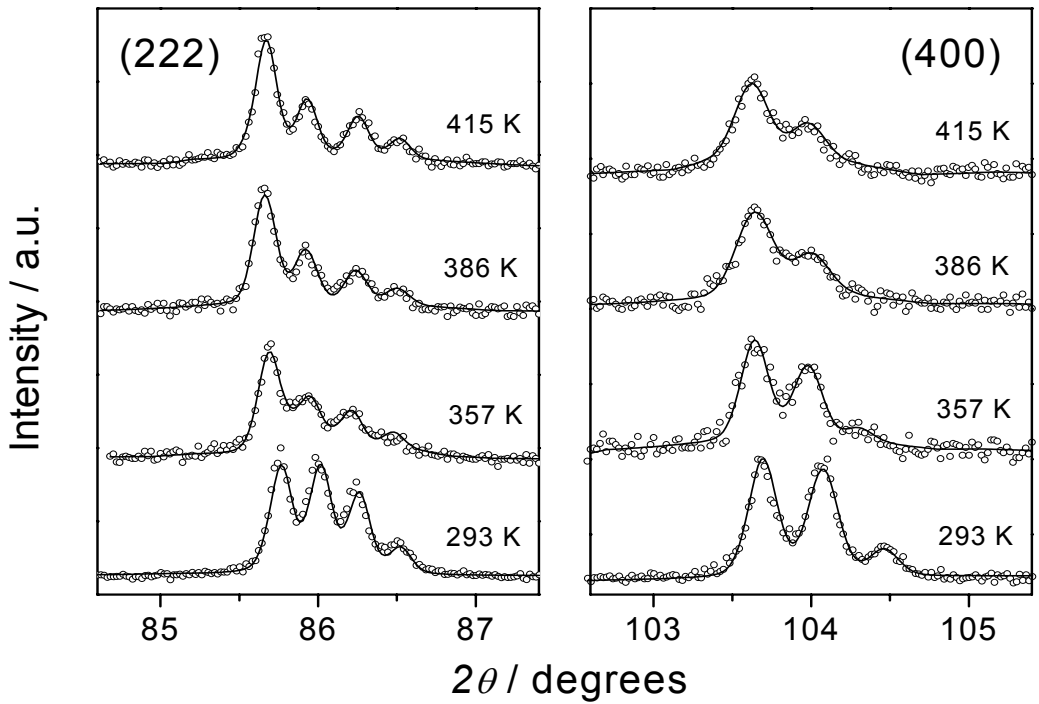

*Fig. 6.* Temperature evolution of the (222) and (400) fundamental reflections for the (1-x)LMT-xLT ceramics with the composition x=0.45.

Figure 5 shows the temperature dependence of $\varepsilon_r$ for these particular compositions. It is seen that the curves of the ceramics with the compositions $x=0.45$ and $0.49$ alter their slope above room temperature suggesting the phase transition, while the ceramics with $x=0.52$ demonstrate linearly decreasing $\varepsilon_r$ over the whole range measured. Lower magnitude of relative permittivity observed in the latter ceramics certainly suggests that these are not a single-phase.

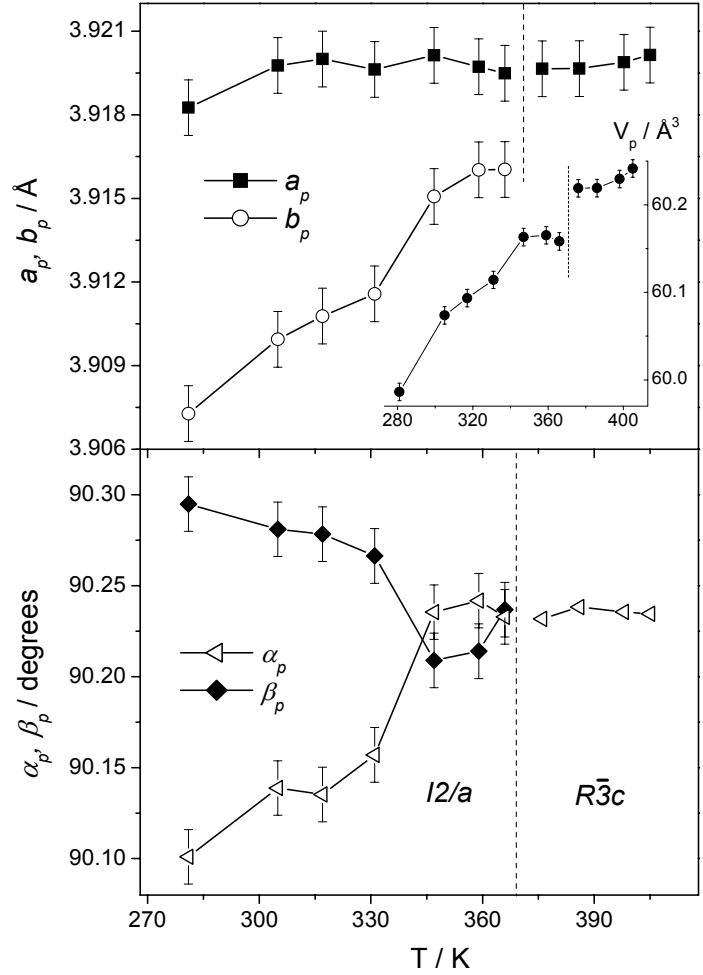

*Fig. 7. Temperature dependence of the primitive perovskite unit cell parameters and cell volume (inset) in (1-x)LMT-xLT (x=0.45).*

To verify the probably structure transformation, the temperature XRD study of the ceramics with $x=0.45$ was performed over the range 300-420 K. As the monoclinic-to-rhombohedral transition was expected, the fundamental reflections (*hkl, h=k=l*) and (*hkl, k=l=0*) being the most sensitive to this type of transformation were examined. The temperature evolution of the (*222*) and (*400*) reflections is shown in Figure 6. As seen, the profiles of the reflections of the composition with $x=0.45$ above 380 K look similar to these of the ceramics with $x=0.52$ (cf. Fig. 2). Indeed, it was unambiguously ascertained by the Rietveld refinement of the XRD spectrum recorded at 405 K that the $I2/a \rightarrow R\bar{3}c$ transition occurs in (1-x)LMT-xLT ($x=0.45$) at about 370 K. The temperature variation of both the parameters of the primitive perovskite unit cell and the unit cell volume testify again a discontinuous type of the transition (Fig. 7). One should pay attention to the features of the $I2/a \rightarrow R\bar{3}c$ crossover when occurring at variation of composition and at variation of temperature. In the former case both $a_p$ and $b_p$ parameters merge into one and its value is between the values of these (Fig. 3). On the other hand, increasing temperature results mainly in altering the value of $b_p$ followed by a jump at the transition, while the value of $a_p$ are near unchanged (Fig. 7). The observed difference in behavior of the parameters at the phase transition seems to be due to an opposite effect of composition and temperature on the unit cell volume of (1-$x$)LMT-$x$LT. While increasing LT content yields a reduce in $V_p$ at a fixed temperature, increasing temperature results in thermal expansion of the crystal lattice of particular composition.

## 4. Conclusions

The single-phase perovskite ceramics were obtained in the (1-$x$)La(Mg$_{1/2}$Ti$_{1/2}$)O$_3$ - $x$La$_{2/3}$TiO$_3$ [(1-$x$)LMT-$x$LT] system at the compositional range $0 \leq x \leq 0.45$. At higher LT content the presence of second phase (La$_2$Ti$_2$O$_7$) was detected. The solubility of end members was found to be corresponded to the actual composition $x=0.52$.

The (1-$x$)LMT-$x$LT solid solution system demonstrates the following structural transformations with the increasing LT content: $P2_1/n$ ($0 \leq x < 0.3$) $\rightarrow$ $Pnma$ ($0.3 \leq x < 0.37$) $\rightarrow$ $Imma$ ($0.37 \leq x < 0.4$) $\rightarrow$ $I2/a$ ($0.4 \leq x \leq 0.49$) $\rightarrow$ $R\bar{3}c$ ($0.49 < x \leq 0.52$). These are related to disappearance of $B$-site ordering (at $x>0.2$) and the change of the oxygen octahedra tilt system from $a^+b^-b^-$ ($x \leq 0.3$), to $a^0b^-b^-$ ($x>0.3$), $a^-b^-b^-$ ($x>0.37$) and finally to $a^-a^-a^-$ ($x>0.49$).

The observed evolution of the structure from the $a^+b^-b^-$ to $a^-a^-a^-$ tilt configuration can be explained by competing the repulsive energy and the Madelung energy associated with these tilt systems. The $a^+b^-b^-$ tilt system allows $A$-site cations to be shifted and is favored in respect of the repulsive energy, while the $a^-a^-a^-$ tilt system requires these cations to remain in their highly symmetrical positions and therefore is preferable with relation to the Madelung energy. The $a^+b^-b^-$ tilt system is the more realizable, when the tilt angles are the larger, i.e. at smaller tolerance factor. In the LMT-LT system, the substitution with LT results in an increase of the tolerance factor. Besides, an amount of the $A$-site vacancies is increased with $x$; thus a number of the shifted cations is decreased and thereby the factors stabilizing the $a^+b^-b^-$ tilt system are suppressed.

While $0 \leq x \leq 0.45$, the values of both relative permittivity and temperature coefficient of the resonant frequency exhibit no drastic changes. In spite of the series of structure phase transitions occurring in (1-$x$)LMT-$x$LT at this range, these do not affect notably the compositional behavior of $\varepsilon_r$ and $\tau_f$. It seems to be due to all these transitions are continuous. However, at higher LT contents the dielectric parameters increase faster with $x$ and $\tau_f$ becomes positive between $x=0.49$ and 0.52, where the discontinuous $I2/a \rightarrow R\bar{3}c$ crossover occurs.

The change of the $\tau_f$ sign from negative to positive for the ceramics with $x \geq 0.45$ results from the change of the $\varepsilon_r$(T) slope at the monoclinic-to-rhombohedral structure transformation, as temperature is increased. Both the temperature-dependent and the composition-dependent $I2/a \rightarrow R\bar{3}c$ transformations are discontinuous, but behaviors of the unit cell parameters at the transition point are different because of an opposite effect of composition and temperature on the unit cell volume of (1-$x$)LMT-$x$LT.

### Acknowledgements

The authors whish to thank the Foundation for Science and Technology (FCT-Portugal) for their financial support through the grants SFRH/BPD/12669/2003 and SFRH/BPD/14988/2004.

**Appendix**

i) Symmetry: monoclinic $P2_1/n$; Refined metric parameters: $a$, $b$, $c$, $\beta$; Metric of primitive perovskite unit cell: $a_p \neq b_p \neq c_p$; $\alpha_p = \gamma_p = 90° \neq \beta_p$.
Vector relationship between monoclinic and primitive perovskite unit cells:
$$\vec{a} = \vec{a}_p + \vec{c}_p; \quad \vec{b} = \vec{a}_p - \vec{c}_p; \quad \vec{c} = 2\vec{b}_p$$
$\vec{a}$, $\vec{b}$, $\vec{c}$ and $\vec{a}_p$, $\vec{b}_p$, $\vec{c}_p$ being lattice vectors of monoclinic and primitive unit cell, respectively.
Metric relationship between monoclinic and primitive perovskite unit cells:
$$a_p = \frac{1}{2}\sqrt{a^2 + b^2 - 2ab\cos\beta}; \quad c_p = \frac{1}{2}\sqrt{a^2 + b^2 + 2ab\cos\beta}; \quad b_p = \frac{c}{2}$$
$$\beta_p = \arccos\left(\frac{a^2 - b^2}{\sqrt{(a^2 + b^2)^2 - 4a^2b^2\cos^2\beta}}\right)$$

ii) Symmetry: orthorhombic $Pnma$; Refined metric parameters: $a$, $b$, $c$; Metric of primitive perovskite unit cell: $a_p = c_p \neq b_p$; $\alpha_p = \gamma_p = 90° \neq \beta_p$.
Vector relationship between orthorhombic and primitive perovskite unit cells:
$$\vec{a} = \vec{a}_p + \vec{c}_p; \quad \vec{b} = 2\vec{b}_p; \quad \vec{c} = \vec{a}_p - \vec{c}_p$$
$\vec{a}$, $\vec{b}$, $\vec{c}$ and $\vec{a}_p$, $\vec{b}_p$, $\vec{c}_p$ being lattice vectors of orthorhombic and primitive unit cell, respectively.
Metric relationship between orthorhombic and primitive perovskite unit cells:
$$a_p = c_p = \frac{1}{2}\sqrt{a^2 + c^2}; \quad b_p = \frac{b}{2}; \quad \beta_p = 2\arctan\left(\frac{a}{c}\right)$$

iii) Symmetry: orthorhombic $Imma$; Refined metric parameters: $a$, $b$, $c$; Metric of primitive perovskite unit cell: $a_p = c_p \neq b_p$; $\alpha_p = \gamma_p = 90° \neq \beta_p$.
Vector relationship between orthorhombic and primitive perovskite unit cells:
$$\vec{a} = 2\vec{b}_p; \quad \vec{b} = \vec{a}_p + \vec{c}_p; \quad \vec{c} = \vec{a}_p - \vec{c}_p$$
$\vec{a}$, $\vec{b}$, $\vec{c}$ and $\vec{a}_p$, $\vec{b}_p$, $\vec{c}_p$ being lattice vectors of orthorhombic and primitive unit cell, respectively.
Metric relationship between orthorhombic and primitive perovskite unit cells:
$$a_p = c_p = \frac{1}{2}\sqrt{b^2 + c^2}; \quad b_p = \frac{a}{2}; \quad \beta_p = 2\arctan\left(\frac{b}{c}\right)$$

iv) Symmetry: monoclinic $I2/a$; Refined metric parameters: $a$, $b$, $c$, $\beta$; Metric of primitive perovskite unit cell: $a_p = c_p \neq b_p$; $\alpha_p = \gamma_p \neq 90° \neq \beta_p$.
Vector relationship between monoclinic and primitive perovskite unit cells:
$$\vec{a} = 2\vec{b}_p; \quad \vec{b} = \vec{a}_p - \vec{c}_p; \quad \vec{c} = \vec{a}_p + \vec{c}_p$$
$\vec{a}$, $\vec{b}$, $\vec{c}$ and $\vec{a}_p$, $\vec{b}_p$, $\vec{c}_p$ being lattice vectors of monoclinic and primitive unit cell, respectively.
Metric relationship between monoclinic and primitive perovskite unit cells:
$$a_p = c_p = \frac{1}{2}\sqrt{b^2 + c^2}; \quad b_p = \frac{a}{2}; \quad \beta_p = 2\arctan\left(\frac{b}{c}\right)$$

$$\alpha_p = \gamma_p = \arccos\left(\cos\beta \cos\left(\arctan\left(\frac{b}{c}\right)\right)\right)$$

v) Symmetry: rhombohedral $R\bar{3}c$; Refined metric parameters in hexagonal setting: $a$, $c$; Metric of primitive perovskite unit cell: $a_p=c_p=b_p$; $\alpha_p=\gamma_p=\beta_p\neq 90^o$.
Vector relationship between hexagonal and primitive perovskite unit cells:

$$\vec{a} = \vec{b}_p - \vec{a}_p; \quad \vec{b} = \vec{c}_p - \vec{b}_p; \quad \vec{c} = 2\vec{a}_p + 2\vec{b}_p + 2\vec{c}_p$$

$\vec{a}$, $\vec{b}$, $\vec{c}$ and $\vec{a}_p$, $\vec{b}_p$, $\vec{c}_p$ being lattice vectors of hexagonal and primitive unit cell, respectively.
Metric relationship between hexagonal and primitive perovskite unit cells:

$$a_p = b_p = c_p = \sqrt{\frac{a^2}{12a^2+c^2}\left(2a+\frac{c^2}{6a}\right)}; \quad \alpha_p = \beta_p = \gamma_p = 2\arccos\left(\sqrt{\frac{3a^2+c^2}{12a^2+c^2}}\right)$$

---